\documentclass{elsart}
\usepackage{amssymb}
\usepackage{graphicx}
\usepackage{epsfig}

\begin{document}


\def\tect{$^{130}$Te }
\def\xect{$^{130}$Xe }
\def\tecv{$^{128}$Te }
\def\pbdd{$^{210}$Pb }
\def\udt{$^{238}$U }
\def\thdt{$^{232}$Th }
\def\tld{$^{208}$Tl }
\def\tectn{$^{130}$Te}
\def\tecvn{$^{128}$Te}
\def\thdtn{$^{232}$Th}
\def\xectn{$^{130}$Xe}
\def\avl{$\langle \lambda \rangle$~}
\def\ave{$\langle \eta \rangle$~}
\def\avm{$\langle g_{\chi\nu} \rangle$~}
\def\mee{$\langle m_{ee} \rangle$~}
\def\mnu{$\langle m_{\nu} \rangle$~}
\def\mmod{$\| \langle m_{ee} \rangle \|$}
\def\mb{$\langle m_{\beta} \rangle$~}
\def\BBz{$\beta\beta(0\nu)$~}
\def\BBm{$\beta\beta(0\nu,\chi)$~}
\def\BBd{$\beta\beta(2\nu)$~}
\def\BB{$\beta\beta$~}
\def\Mz{$|M_{0\nu}|$~}
\def\Md{$|M_{2\nu}|$~}
\def\Tz{$T^{0\nu}_{1/2}$~}
\def\Td{$T^{2\nu}_{1/2}$~}
\def\Tm{$T^{0\nu\,\chi}_{1/2}$~}
\def\ca{$\sim$}
\def\dca{$\approx$}
\def\dot{$\cdot$}
\def\pom{$\pm$ }
\def\gm{$\gamma$}
\def\ne{$\neq$}
\def\teod{TeO$_2$~}
\def\teodn{TeO$_2$}
\def\be{\begin{equation}}
\def\ee{\end{equation}}

\begin{frontmatter}

\title{A Calorimetric Search on Double Beta Decay of \tect }

\author[label1] {C. Arnaboldi},
\author[label1]{C. Brofferio},
\author[label3]{C. Bucci},
\author[label1]{S. Capelli}, 
\author[label1]{O. Cremonesi}, 
\author[label1]{E. Fiorini\thanksref{Spokesman}}, 
\author[label2]{A. Giuliani},
\author[label1]{A. Nucciotti},
\author[label1]{M. Pavan},
\author[label2]{M. Pedretti}, 
\author[label1]{G. Pessina}, 
\author[label1]{S. Pirro},
\author[label3]{C. Pobes\thanksref{CEE}},
\author[label1]{E. Previtali}, 
\author[label1]{M. Sisti},
\author[label1]{M. Vanzini}.

\address[label1]{Dipartimento di Fisica dell'Universit\`{a} di 
Milano-Bicocca e Sezione di Milano dell'INFN, Milan I-20126,
Italy} 

\address[label2]{Dipartimento di Scienze Chimiche, Fisiche e
Matematiche dell'Universit\`{a} dell'Insubria e Sezione di Milano
dell'INFN, Como I-22100, Italy }

\address[label3]{Laboratori Nazionali del Gran Sasso, I-67010,
Assergi (L'Aquila), Italy}

\thanks[Spokesman]{c.a.: Ettore Fiorini, Piazza della Scienza 3, 20126 Milano (Italy) Tel +39-02-64482424  Fax +39-02-64482463  e-mail ettore.fiorini@mi.infn.it}              
\thanks[CEE]{CEE fellow  in the Network on Cryogenic Detectors, under contract FMRXCT980167, presently at the Lab. De Fisica Nuclear y Altas Energias, University of Zaragoza (Spain)}

\begin{abstract}
We report on the final results of a series of experiments on double beta decay of  
\tect carried out with an array of twenty cryogenic detectors. The set-up is made
with crystals of \teod with a total mass of 6.8 kg, the largest operating one for 
a cryogenic experiment. Four crystals are made with isotopically enriched 
materials: two in \tecv and two others in \tect. The remaining ones are made with 
natural tellurium, which contains 31.7~\% and 33.8~\%  \tecv and \tectn, 
respectively. The array was run under a heavy shield in the Gran Sasso Underground 
Laboratory at a depth of about 3500 m.w.e. By recording the pulses of each
detector in anticoincidence with the others a lower limit of $2.1 
\times 10^{23}$ years has been obtained at the 90~\% C.L. on the lifetime for 
neutrinoless double beta decay of \tect.\\ 
In terms of effective neutrino mass this 
leads to the most restrictive limit in direct experiments, after those obtained with Ge 
diodes. Limits on other lepton violating decays of \tect  and on the neutrinoless 
double beta decay of \tecv to the ground state of $^{128}$Xe are also reported and 
discussed. An indication is presented for the two neutrino double beta decay 
of \tect. Some consequences of the present results in the interpretation of 
geochemical experiments are discussed.
\end{abstract}

\end{frontmatter}

\section{Introduction}

Double beta decay (DBD), in its two negatron channel, consists in a rare
transition from the nucleus (A,Z) to its isobar (A,Z+2) with the emission of two electrons. 
It can be searched for when the transition from (A,Z) to (A,Z+1) is 
energetically forbidden or at least strongly hindered by a large change of the 
spin-parity state. 
This process can occur into various channels:
\be
(A,Z) \to (A,Z+2) + 2 e^-  + 2\overline{\nu}_e 	
\label{eq:bb2nu}
\ee
\be
(A,Z) \to (A,Z+2) + 2 e^-  +  ( N)\chi \;\; [N=1,2,...]
\label{eq:bbmaj}
\ee
\be
(A,Z) \to (A,Z+2) + 2 e^-
\label{eq:bb0nu}
\ee
where  $\chi$  is a massless Goldstone boson named Majoron.
All three double beta processes can also occur to  excited  states of the 
daughter nucleus with a consequent lower ``effective'' transition energy.
The lepton conserving process of two neutrino DBD, which is allowed by the 
Standard Model, has been revealed in 10 nuclei for the transition to the ground 
state and in one case also for the transition to an excited level of the 
daughter nucleus ~\cite{Elliott02,Cremonesi02,Tretyak02,Ejiri01,Ejiri02}. Both processes (\ref{eq:bbmaj}) and (\ref{eq:bb0nu}) violate the lepton number conservation and are forbidden by the Standard Model. The third, normally called ``neutrinoless DBD'' 
is experimentally very appealing, since it could 
be revealed by a peak corresponding to the total transition energy, since the 
nuclear recoiling energy is negligible. 
No evidence has been claimed so far for the leptonic violating channels 
(\ref{eq:bbmaj}) and (\ref{eq:bb0nu}), with the exception of an alleged evidence for neutrinoless 
DBD reported by Klapdor-Kleingrothaus  et al.~\cite{Klapdor01,Harney02} but confuted by other 
authors~\cite{Feruglio02,Aalseth02}.\\Experiments for neutrinoless DBD 
represent a powerful tool to search for a finite ``effective'' neutrino mass  and/or 
for the presence of right handed currents in the weak interaction amplitude. 
A non-zero \mnu has been  suggested 
~\cite{Ejiri02,Raffelt02,Barger02,Hong02,Haba02,Zhi02,Pascoli02,Zdesenko02,Fiorini02} 
by the recent evidence of neutrino oscillations 
~\cite{Ahmad02,Fukuda00,Fukuda01,Nishikawa02}.\\An indirect approach to search for DBD consists in radiochemical 
experiments~\cite{Turkevich92} where the material containing the parent nucleus 
is stored underground for a long period and later searched for the decay of the 
radioactive daughter nucleus.
Geochemical experiments~\cite{Manuel86,Kirsten86,Bernatowicz93,Takaoka96,Kawashima93} 
consist in the search for an abnormal abundance of the (A,Z+2) isotope 
extracted from a geological old rock containing a substantial amount of the nucleus (A,Z). 
These experiments are very sensitive, due to the long ``exposure time'', 
but, like the radiochemical ones, indicate only the presence of the 
daughter nucleus and cannot therefore discriminate between lepton conserving and non conserving 
processes or between decays to the ground or excited states of the daughter 
nucleus. Some indication for DBD~\cite{Kawashima93} of $^{96}$Zr and  definite 
evidence for DBD of $^{82}$Se, \tecv and \tect  have been 
reported~\cite{Elliott02,Cremonesi02,Tretyak02,Ejiri01,Ejiri02,Manuel86,Kirsten86,Bernatowicz93,Takaoka96}.\\Direct 
experiments are based on two different methods. In the 
source$\neq$detector approach a double beta active material is inserted, 
normally in form of thin sheets, in a suitable detector. In the source=detector 
or ``calorimetric'' experiments~\cite{Fiorini60} the detector itself is made by a 
material containing the double beta active nucleus. 
The use of cryogenic detectors to search for DBD has been suggested in 
1984~\cite{Fiorini84}. These detectors~\cite{Twerenbold96,LTD01} are based on 
the peculiar property of the heat capacity of  
diamagnetic and dielectric crystals which, at low temperature, is proportional  
to the cube of the ratio between the operating and Debye temperatures. 
As a consequence in a 
cryogenic set-up this capacity can become so small that even the tiny energy 
released by a particle in form of heat can be revealed by the temperature 
increase of the absorber by means of a suitable thermal sensor.  
Unlike conventional detectors, the 
cryogenic ones offer a wide choice of  DBD candidates, the only requirement 
being that the candidate nucleus be part of a compound which can be grown in 
the form of a crystal with reasonable thermal and mechanical properties.
\tect looks an excellent candidate to search for DBD due to its high transition 
energy (2528.8 \pom 1.3 keV)~\cite{Dyck90}, and large isotopic abundance 
(33.8~\%)\cite{Firestone98} which allows to perform a sensitive experiment with 
natural tellurium. In addition, the expected signal at 2528.8 keV is in an 
energy region between the peak and the Compton edge of the \tld \gm-rays at 
2615 keV, which generally dominates the background in this high energy region.\\Results on neutrinoless 
double beta decay of \tect have been already obtained 
with one~\cite{Alessandrello94}, four~\cite{Alessandrello96} and 
eight~\cite{Alessandrello98}  detectors made by 340 g crystals of \teodn. The 
first operation and the preliminary results of an array of 20 crystals of 
natural \teod of 340 g each, operated in coincidence and anticoincidence, has 
been presented previously~\cite{Alessandrello00}. 
We report here new results on two neutrino and neutrinoless DBD, based on 
substantial improvements of the detector and  larger statistics. 

\section{Experimental details}

The array consists in a tower with five planes of 4 detectors each, operating in a dilution refrigerator 
in the Gran Sasso Underground Laboratory~\cite{Bettini01}. The twenty absorbers are crystals of \teod of 3x3x6 cm$^3$ volume with a total active mass of about 6.8 kg, the largest in any cryogenic experiment. Sixteen of these crystals are made of natural telluride. Two contain tellurium isotopically enriched at 82.3~\% in \tecv and other two at 75.0~\% in \tect. By mass spectrometer measurements we found that these enrichments are lower than the original ones (94.7~\% and 92.8~\%, respectively). This is due to the process of crystallization which is more complex when isotopically enriched powder is used, and requires seeds of natural telluride. 
\begin{figure}[bt]
    \begin{minipage}[c]{0.95\textwidth}%
      \centering\includegraphics[height=8cm]{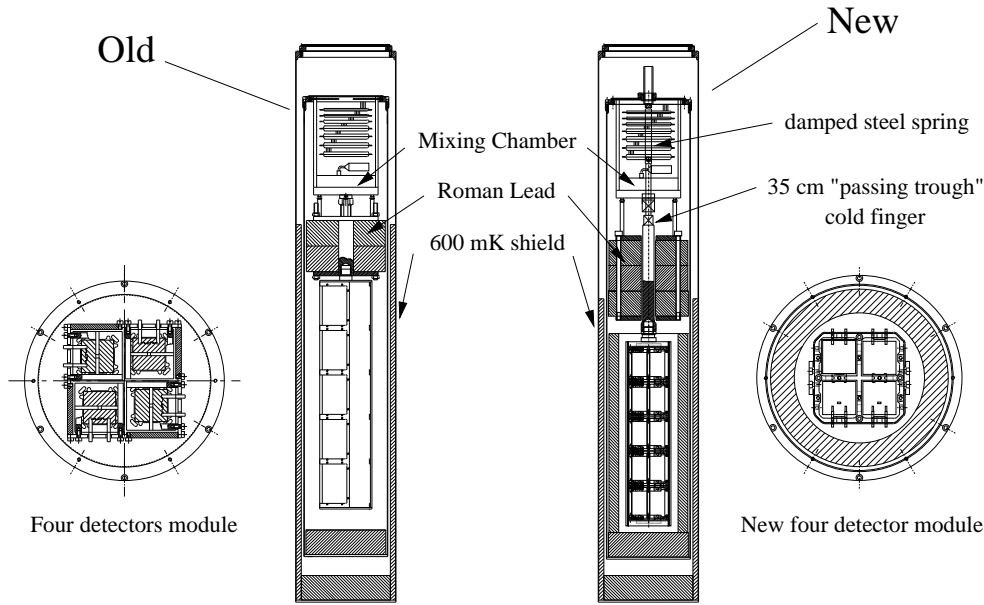}%
    \end{minipage}%
\caption{Scheme of the array of twenty detectors.}
\label{fig:towers}
\end{figure}
The temperature sensors are Neutron Transmutation Doped (NTD) Ge thermistors thermally coupled to each crystal. 
They were specifically prepared in order to present similar thermal performance.  A resistor of 100 - 200 k$\Omega$, 
realized with a heavily doped meander implanted on a 1 mm$^3$ silicon chip, was attached to each absorber and acted 
as a heater to calibrate and stabilize the gain of the bolometer~\cite{Alessandrello98a}.
 The tower is connected via an OFHC copper cold finger to the mixing chamber of a dilution refrigerator specially 
 constructed with previously tested low radioactivity materials. The entire set-up is shielded with two layers of 
 lead of 10 cm minimum thickness each. The outer one is made of common low radioactivity lead, the inner of special 
 lead with a contamination of 16 \pom 4 Bq/kg in \pbdd. The electrolytic copper of the refrigerator thermal shields 
 provides an additional shield of 2 cm minimum thickness.\\The new results reported here refer to runs carried out with two different configurations (Fig. \ref{fig:towers}). 
In the former one the intrinsic radioactive contamination of the dilution unit materials (e.g. from silver and 
stainless steel) is shielded by a layer of 10 cm Roman lead (\pbdd activity $<$~4 mBq/kg~\cite{Alessandrello98b}), 
framed inside the cryostat immediately above the tower of the array. The refrigerator is surrounded by a Plexiglas 
anti-radon box fluxed with clean N$_2$ from a liquid nitrogen evaporator, and by a Faraday cage to eliminate 
electromagnetic interference.
In the latter configuration the following improvements have been implemented:
\begin{description}
\item[a.] all crystals have been thoroughly lapped with previously tested low radioactivity powder to reduce the 
surface contamination introduced by the original production process in China. All these operations and the final 
mounting of the tower were carried out in a clean box.
\item[b.] a more compact assembling of the crystals has been adopted. This has allowed us to add an internal lateral 
shield of Roman lead of 2 cm minimum thickness and to increase by 5 cm the thickness of the lead shield between the 
mixing chamber and the detector.
\item[c.] a spring to which the tower is hanged has been added to reduce vibrations.
\end{description}
The front-end electronics of each detector is located at room temperature. It consists of a differential voltage 
sensitive preamplifier followed by a second stage and an antialiasing 
filter~\cite{Alessandrello97b,Alessandrello00b,Pessina00}. The differential configuration has been adopted in order to minimize signal cross talk and microphonic 
noise  coming from the connecting wires. Precautions have been taken to suppress any possible effect coming from any 
room temperature drift~\cite{Alessandrello97b} and main supply instability~\cite{Pessina99}. 
A pair of load resistors serves to bias each bolometer in a symmetric way. They are located at room temperature, 
close to the preamplifier, and consist of metal films (Micro-Ohm) with a value of 30 G$\Omega$ each. Their 
manufacturing process and large value have been chosen in order to maintain as low as possible their thermal 
and low frequency noise contribution~\cite{Arnaboldi02}. 
All the necessary settings for the front-end and the biasing system are programmed remotely via computer, 
in order to allow the optimization of the overall dynamic performance separately for each 
detector~\cite{Alessandrello00b}.\\All other details, which are common to both set-ups have been reported previously~\cite{Alessandrello00}.
In both cases the array was cooled down to temperatures around 8 mK with a temperature spread of  \ca 1 mK 
among the different detectors.
The detectors were calibrated by a combined radioactive source of \udt and  \thdtn. Their FWHM energy 
resolutions at the 2615 keV \tld line range from 5 to 15 keV.

\section{Experimental results}

The two set-ups have been operated for effective running times of \ca 31,508 and
\ca 5,690 hours\dot kg, 
respectively. Sum spectra have been obtained both with no anticoincidence cut and  by operating 
each detector in anticoincidence with all the others.  In all these spectra the main lines due to 
the natural activity of the  \thdt and \udt chains,  of $^{40}$K and the lines at 1173 and at 1332 keV 
due to cosmogenic $^{60}$Co are present.\\No peak appears in the region of neutrinoless DBD of  \tectn, 
where the rates are, respectively, of 0.59 \pom 0.06 and 0.33 \pom 0.11 counts 
keV$^{-1}$ kg$^{-1}$ year$^{-1}$ for the former and latter run, when operated in anticoincidence. 
No peak also appears at the energies  corresponding to neutrinoless DBD of \tect  to 
excited levels of \xectn, and at the energy of 867 keV corresponding to neutrinoless DBD of \tecvn.\\
\begin{figure}[bt]
    \begin{minipage}[c]{0.95\textwidth}%
      \centering\includegraphics[width=1\textwidth]{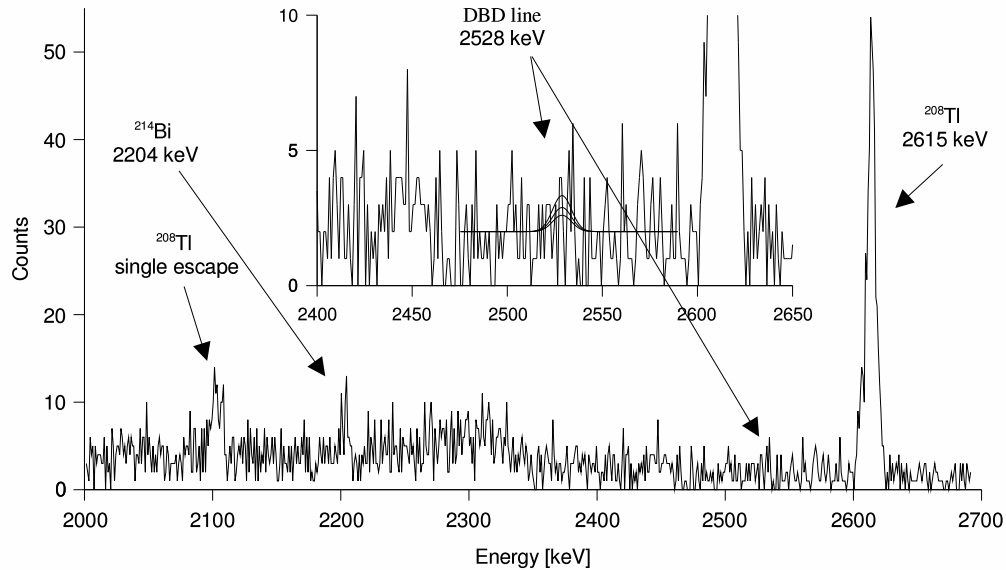}%
    \end{minipage}%
\caption{Total spectrum (in anticoincidence) in the region of neutrinoless DBD obtained with the twenty  crystal array. The solid curves represent the best fit (lowest curve) and the 68~\% and 90~\% C.L. excluded signals.}
\label{fig:sbb0n}
\end{figure}

The sum of the spectra obtained in anticoincidence in the two runs  in the  region 
above 2000 keV is shown in Fig. \ref{fig:sbb0n}. It corresponds  to \ca 3.56 kg x year of \teod  and 
to \ca 0.98  kg x year of \tect. The clear peaks corresponding to the lines at 2104 keV (single escape 
of the 2615 keV \tld line), at 2204 keV ($^{214}$Bi) and at 2615 keV (\tld), confirm the reproducibility 
of the array during both runs.\\ We would like to note that in the spectrum of the second run both the peaks and
the continuum were decreased by a factor of about two with respect to the first run. 
However, according to the results obtained from the background analysis and supported by the Monte Carlo simulation of the detector,
these similar reduction factors are due to two different effects.
 While the peaks were due to sources outside the detector and were consequentely reduced by
the increased shield of lead, the continuum was mainly reduced by the lapping and cleaning of the
crystals and copper frame. This 
confirms that the origin of the background at the energy corresponding to neutrinoless DBD is mainly due to
surface contamination. 

Fit parameters and 90~\% C.L. limits for the various decay processes were evaluated with a maximum likelihood procedure. 
Assuming a Poisson statistics for the binned data the fit procedure was formulated in terms of the likelihood 
chisquare ($\chi^2_L$) \cite{Baker84}: 
\begin{equation}
\chi^2_L=2\sum_i(y_i-n_i+n_i \, ln(n_i/y_i))
\end{equation}
where $n_i$ is the number of events in the i-th spectrum bin and $y_i$ the number of events predicted by the fit model. 
Fit parameters were estimated by minimizing $\chi^2_L$, while limits were obtained, after proper renormalization, just 
considering the $\chi^2_L$ distribution in the physical region \cite{Barnett96}. A global fit procedure based on the 
minimization of $\chi^2_T=\sum_i\chi^2_i$ was adopted to combine different measurements. Similar results were obtained 
following the approach proposed by G.J. Feldman and R.D.~Cousins~\cite{Feldman98}, 
suggested by the Particle data Group~\cite{Groom00}.

\begin{table}[htb]
\caption{Half lifetime limits (90~\% C.L.) on lepton violating and conserving channels. E$_0$ is the energy at which T$_{1/2}$ 
was obtained while \emph{a.c.} means anticoincidence between different detectors.}
\label{tab:limits}
\begin{tabular}{@{}ccccccc}
\hline\hline
Isotope & Transition & Used Spectra & Cuts & E$_0$ & Efficiency & T$_{1/2}$\\
& & & & (keV) & (\%) & (years)\\
\hline
\tect & $0\nu: 0^+ \to 0^+$ & All spectra & a.c. & 2528.8 & 84.5 & $> 2.1 \times 10^{23}$\\
\tect & $0\nu: 0^{+*} \to 0^+$ & All spectra & none & 1992.8 & 7.9 & $> 3.1 \times 10^{22}$\\
\tect & $0\nu: 0^{+*} \to 2^+$ & All spectra & none & 1992.8 & 37.5 & $> 1.4 \times 10^{23}$\\
\tect & $2\nu: 0^+ \to 0^+$ & \tect crystals & a.c. &  &   & $> 3.8 \times 10^{20}$\\
\tect & $1\chi: 0^+ \to 0^+$ & \tect crystals & a.c. &  &  & $> 2.2 \times 10^{21}$\\
\tect & $2\chi: 0^+ \to 0^+$ & \tect crystals & a.c. &  &  & $> 0.9 \times 10^{21}$\\
\tecv & $0\nu: 0^+ \to 0^+$ & All spectra & a.c. & 867.2 & 97.9 & $> 1.1 \times 10^{23}$\\
\hline\hline
\end{tabular}
\end{table}
Efficiencies and limits for the various decay channels were estimated as follows:
\begin{description}
\item[a.] The spectra used to evaluate neutrinoless DBD to the ground level were obtained by considering only 
events which are, as expected, contained in a single detector. By rejecting all multiple events we have achieved 
a reduction of the background in the neutrinoless DBD region of \ca 25~\%. 
\item[b.] A  neutrinoless DBD to a $0^{+*}$ intermediate state at 1072 keV has been suggested in 
ref. \cite{Aunola96}. It would be followed by the decay of this state to the 536 keV $2^{+}$ level, 
followed by the emission of a second  \gm-ray. The limit for this process can be evaluated in various ways. 
We found that the more restrictive limit could be obtained from the spectrum without anticoincidence, 
assuming that the same crystal absorbs the two electrons and the first \gm-ray, while the 536 keV one 
escapes from it. Efficiency and limit have been evaluated accordingly. 
\item[c.] Efficiency and rate for the neutrinoless $0^+ \to 2^+$  DBD to the 536 keV state have been also 
evaluated from the spectrum without anticoincidence and assuming the escape of the 536 keV \gm-ray.
\item[d.] The constraints for two neutrino and majoron decays have been estimated from the spectrum obtained with 
each \tectn{O$_2$} crystal in anticoincidence with the other crystals of the array. The corresponding limits have 
been conservatively obtained by evaluating the maximum areas for which the expected spectra do not exceed the 
background spectra and their fluctuations at any energy. The detection efficiency for these processes is near 
to one.
\end{description}

\section{Lepton non conserving double beta decay}

The constraints  on the effective neutrino mass \mnu, on the right handed current parameters \avl 
and  \ave and on the coupling majoron parameter \avm suffer from uncertainties in the calculation of the 
nuclear matrix elements~\cite{Faessler98,Barbero99,Engel88,Engel89,Muto89,Suhonen91,Tomoda91,Stoica01,Faessler01}. 
The limits on the various channels of neutrinoless DBD of \tect are reported in Table \ref{tab:lncp} 
on the basis of various QRPA calculations, since the shell model does not seem
appropriate for heavy nuclei ~\cite{Faessler98}. Taking into account theoretical uncertainties we obtain from our data constraints in the ranges (1.1-2.6) eV;
   $(1.6-2.4)\times 10^{-6}$ and $(0.9-5.3)\times 10^{-8}$ for the values of effective neutrino mass, and 
   the two right handed parameters  $\lambda$  and $\eta$ respectively. Our limit on  \mnu  appears to be 
the most restrictive one among those obtained with direct methods after those on $^{76}$Ge (0.35-1.4 eV).

\begin{table}[htb]
\caption{Limits on the lepton non-conserving parameters from this experiment. For each parameter, limits 
are abtained assuming vanishing the others.}
\label{tab:lncp}
\begin{tabular}{@{}lcccc}
\hline\hline
 Ref.& \mnu & \avl$\times 10^{-6}$& \ave$\times 10^{-8}$ & \avm$\times 10^{-5}$ \\
\hline

Engel et al. (1988)\cite{Engel88}& 1.8 & & & 28 \\
Engel et al. (1989)\cite{Engel89}& 1.1 & & & 17 \\
Muto et al. (1989)\cite{Muto89}& 1.5 & 2.1 & 1.4 & 24 \\
Suhonen et al. (1991)\cite{Suhonen91}& 2.0 & 2.1 & 5.3 & 31 \\
Tomoda et al. (1991)\cite{Tomoda91}& 1.6 & 2.4 & 1.6 & 25 \\
Faessler et al. (1998)\cite{Faessler98}& 2.1 &  &  & 33 \\
Barbero et al. (1999)\cite{Barbero99}& 1.3 & 1.6 & 0.9 & \\
Klapdor-Kleingrothaus and& 2.2-2.3 &  &  & \\
 Stoica(2001)\cite{Stoica01}& &  &  & \\
Faessler and Simkovic (2001)\cite{Faessler01}& 2.6 & 2.4 & 1.8 &  \\
\hline\hline
\end{tabular}
\end{table}
Our exclusive limit on the half lifetime for neutrinoless DBD of \tecv is the best in direct experiments,
 but less constraining than those extracted (for the same nucleus) from the inclusive limits of geochemical
  experiments~\cite{Manuel86,Kirsten86,Bernatowicz93,Takaoka96}. 
  
\section{Lepton conserving double beta decay}

\begin{figure}[bt]
    \begin{minipage}[c]{0.95\textwidth}%
      \centering\includegraphics[height=8cm]{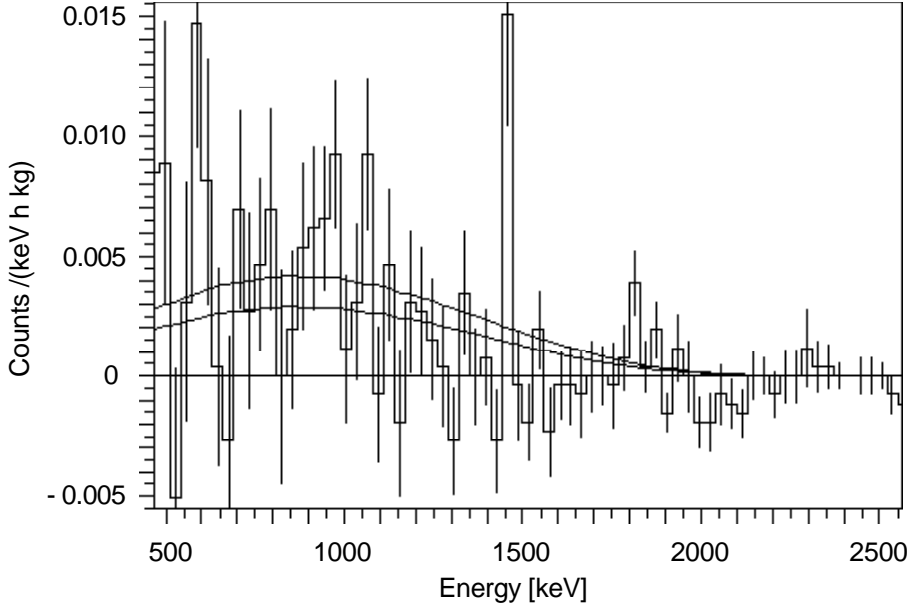}%
    \end{minipage}%
\caption{Total difference spectrum between \tect and \tecv detectors (no background subtraction). The solid curves represent the best fit (lowest curve) and the 90~\% C.L. excluded signal (Tab. \ref{tab:limits})}
\label{fig:sbb2n}
\end{figure}

The 90~\% C.L. limit for two neutrino DBD of \tect reported in Table \ref{tab:limits} already excludes a relevant loss of the daughter isotope in geochemical experiments, which has been  considered by O.K. Manuel~\cite{Manuel86}. 
We have also attempted to obtain an evaluation of the rate for this lepton conserving process by analyzing the difference between the sum of the two spectra of the crystals isotopically enriched in \tect and the sum of those  of the crystals enriched in \tecv. (Fig. \ref{fig:sbb2n}).
These differences are positive in the region of two neutrino DBD with an excess of $269\pm60$ counts corresponding to $T_{1/2}=(6.1\pm 1.4\,stat.)\times 10^{20}$ years.
However a precise evaluation of the half life is not straightforward because the background differs among the four enriched crystals. In particular different rates are observed for the 1460 keV gamma line of $^{40}$K and for the alpha lines in the 4-6 MeV energy region.
With the aid of a Monte Carlo simulation of the possible background 
contributions we were able to restrict the possible sources responsible 
of these lines. The 1460 keV gamma peak appears to be due to an 
accidental $^{40}$K contamination localized on the bottom surface of 
the detector copper holder, while the alpha lines are clearly due to 
$^{238}$U and  $^{232}$Th surface contaminations of the crystals. A similar 
- although lower - surface contamination is observed also 
for all the natural crystals.
Taking into account the different rates observed in the lines of our 
isotopically enriched detectors we extrapolated the contribution 
to the $^{130}$Te - $^{128}$Te spectra produced by such contaminations 
in the two neutrino DBD region. The maximum expected difference due 
to $^{238}$U and $^{232}$Th contaminations
is negative  and corresponds to 380 conts, which should therefore be 
added to the actually found signal. On the contrary the corresponding 
maximum  expected  difference due to background from $^{40}$K
(86 counts) is  positive and should therefore be subtracted.  Slightly 
different values are obtained by varying the contaminations
location. In all the cases however negative values are obtained for 
the $^{238}$U and $^{232}$Th contribution 
(responsible of the differences in the alpha counting rates)
and positive for the corresponding  $^{40}$K ones.
We conclude that the difference in the crystal background rates cannot 
account for the two neutrino DBD effect, but introduce a large systematics 
in the half life time evaluation.
By assuming the above quoted background rates as the maximum possible 
contribution to our systematic error, our final result is
 $T_{1/2}=(6.1\pm 1.4\, stat.^{+2.9}_{-3.5}\,sys.)\times 10^{20}$ years.
 This value, while in agreement with most geochemically obtained results, 
looks somewhat higher than most of those predicted theoretically~\cite{Tretyak02}. 
The already running NEMO 3 experiment~\cite{Piquemal01}, as well as the an improved search to be carried out 
with the larger CUORICINO array~\cite{Alessandrello00c} being mounted in the Gran Sasso laboratory, will 
allow to reduce soon the present uncertainty.

\section{Comparison with geochemical experiments}

Geochemical experiments simply indicate the presence of the (A,Z+2) nucleus in the sample containing (A,Z). 
The quoted rates for \tect refer therefore to the sum of all possible transitions to the ground or  
excited levels, with or without the emission of  neutrinos or  majorons. Our indication for two neutrino DBD 
of \tect is in qualitative agreement with the inclusive rates indicated in most geochemical experiments.\\On the other side 
our exclusive limits on individual neutrinoless DBD processes allow to constrain their 
contribution to the inclusive values found geochemically. In particular our limit on neutrinoless DBD 
of \tect excludes contributions of 0.3 to 1.3~\% to the overall rate in the two extreme results of 
Manuel et al~\cite{Manuel86} and Bernatowicz et al.~\cite{Bernatowicz93}, respectively. 
Our limit on neutrinoless DBD to a hypothetical 0$^+$ excited state at 1072 keV~\cite{Aunola96} indicates that 
this process cannot be responsible for more than 2 and 9~\% of the geochemically obtained rates in the two 
extreme results reported above. The corresponding maximal contributions of neutrinoless DBD to the 2$^+$ 
excited state of \xect at 536 keV are 0.5 and 2~\%, respectively. By applying the same procedure to our limit 
on DBD mediated by a majoron we achieve upper limits of  33 and 87~\%. 

\section{Conclusions}

No evidence is found in this experiment for neutrinoless DBD of \tect with a 90~\% C.L. lower limit of  
$2.1 \times 10^{23}$ years. This corresponds to an upper limit on the effective neutrino mass \mnu ranging 
from 1.1 to 2.6 eV, on the basis of the various evaluations of the nuclear matrix elements. This, as well 
as the limits on the values of the contributions of right handed currents and on  the possible majoron coupling, 
are the most constraining ones in direct experiments after those obtained with Ge diodes.\\We have an indication for two neutrino double beta decay of \tect to the ground state of  \xect which roughly 
confirms the positive evidence found in geochemical experiments. Our experiment also shows that the neutrinoless 
channels do not account for relevant contributions to the overall DBD rate of \tect measured in geochemical 
experiments.\\The peculiarly large isotopic abundance of  \tect allows the use of natural telluride in large scale searches 
for neutrinoless DBD. An experiment, named CUORE (for Cryogenic Underground Observatory for Rare Events) 
made by 1000 crystals of \teod with a total mass of almost 800 kg is being studied~\cite{Alessandrello00c} and 
a smaller array, named CUORICINO, totalling  about 40 kg of active \teod, is being constructed~\cite{Brofferio02}. 


Thanks are due to the Laboratori Nazionali del Gran Sasso for generous hospitality and to C.~Callegaro, R.~Gaigher,
 S.~Parmeggiano, M.~Perego, B.~Romualdi and  A.~Rotilio and to our student P.~Gorla for continuous and constructive
  help in various stages of this experiment. We also gratefully acknowledge contributions of  A.~Alessandrello and L.~Zanotti in the first stage of our cryogenic activity in Gran Sasso. Thanks are also due to Till~Kirsten 
  for enlightening discussions on the results of geochemical experiments. We are indebted to Enzo Palmieri and to the
  Laboratori Nazionali di Legnaro for the surface treatment of all copper components and essential advices.\\This 
  experiment has been supported in part by the Commission of European Communities under contract FMRX-CT98-03167

\end{document}